\documentclass[11pt]{article}
\usepackage{bm}
\usepackage{a4wide}
\usepackage{amsmath,amssymb}

\DeclareMathOperator{\diag}{diag}
\DeclareMathOperator{\blockdiag}{blockdiag}

\begin{document}
\title{Orbifold Family Unification}

\author{\Large
		Yoshiharu \textsc{Kawamura}$^1$\thanks{
			E-mail: \tt haru@azusa.shinshu-u.ac.jp}\and\Large
		Teppei \textsc{Kinami}$^1$\thanks{
			E-mail: \tt s06t303@shinshu-u.ac.jp}\and\Large
		Kin-ya \textsc{Oda}$^2$\thanks{
			E-mail: \tt odakin@riken.jp}\bigskip\\
	$^1$\it Department of Physics, Shinshu University \\
	    \it Matsumoto 390-8621, Japan \smallskip\\
	$^2$\it Theoretical Physics Laboratory, RIKEN\\
	    \it Saitama 351-0198, Japan\\
	\\
	}


\maketitle
\begin{abstract}
\noindent \normalsize We study the possibility of complete family unification in higher-dimensional space-time.
Three families of matters in $SU(5)$ grand unified theory are derived from a single bulk multiplet of $SU(N)$ gauge group 
($N \ge 9$) in the framework of $S^1/Z_2$ orbifold models.
In the case of the direct orbifold breaking down to the standard model gauge group, there are models in which
bulk fields from a single multiplet and a few brane fields compose three families of quarks and leptons. 
\end{abstract}

\vfill

\rightline{RIKEN-TH-94}

\newpage

\section{Introduction}

{\it Unification} is the paradigm in which physical laws and substances have been successfully organized.
In quantum field theory, the unification of substances is realized by symmetry principle, i.e.,
elementary particles are classified by irreducible representations of some transformation group.
Interactions among elementary constituents are also determined by the symmetry.
The {\it grand unification} offers unification of force and (partial) unification of 
quarks and leptons in each family~\cite{G&G,LR}.
Still the origin of the family replication has been a big riddle.
The {\it family unification} based on larger symmetry group provides a possible solution~\cite{R,GF,K&Y}.
Especially, we refer to the unification of all the three families of quarks and leptons within a single representation as the {\it complete family unification}.

In the four-dimensional Minkowski space-time, we encounter difficulty in (complete) family unification
because of extra fields such as `mirror particles' existing in the higher-dimensional representation.
The mirror particles are particles with opposite quantum numbers under the standard model (SM) gauge group.
If the idea of (complete) family unification is to be realized in nature, extra particles must disappear 
from the low-energy spectrum around the weak scale.
Several interesting mechanisms have been proposed to get rid of the unwelcomed particles.
One is to confine extra particles at a high-energy scale by some strong interaction~\cite{GR&S}.
Another possibility is to reduce symmetries and substances using extra dimensions, 
as originally discussed in superstring theory~\cite{CHS&W,DHV&W}.

Higher-dimensional grand unified theories (GUTs) on an orbifold possess attractive features as a realistic model.
The triplet-doublet splitting of Higgs multiplets is elegantly realized in supersymmetric (SUSY) $SU(5)$ GUT 
in five dimensions~\cite{K,H&N}.
In the model, four-dimensional chiral fermions are generated through the dimensional reduction 
where a part of zero modes can be projected out 
by orbifolding, i.e., by non-trivial boundary conditions (BCs) concerning extra dimensions on bulk fields.
Therefore we expect that all the extra particles 
plaguing the family unification models can possibly be eliminated from the spectrum 
in the framework of orbifold GUTs and that the idea of complete family unification can be realized.\footnote{
	The possibility that one might achieve the complete family unification utilizing an orbifold has been also suggested in the earlier reference~\cite{BB&K} in a different context.
	In Ref.~\cite{Watari:2002tf}, three families have been derived from a combination of a bulk gauge multiplet and a few brane fields.
	In Ref.~\cite{Chaichian:2001fs}, they have been realized as composite fields.}

In this paper, we pursue this possibility of complete family unification in higher-dimensions.
For $SU(N)$ gauge theory on the orbifold $S^1/Z_2$, we investigate whether or not
three families are derived from a single bulk multiplet in two types of orbifold breaking, namely to $SU(5)$ and to the SM gauge groups.
The validity of our analysis is discussed from the viewpoint of equivalence classes of the BCs on $S^1/Z_2$.

The contents of this paper are as follows.
In section 2, we review and provide general arguments on the orbifold breaking on $S^1/Z_2$.
In section 3, we investigate unification of quarks and leptons using $SU(N)$ orbifold GUTs
and discuss the validity of our analysis. 
Section 4 is devoted to conclusions and discussions.

\section{$S^1/Z_2$ orbifold breaking}

\subsection{Preparations}

First we review the argument in~\cite{HHH&K}.
The reader familiar with the $S^1/Z_2$ orbifold symmetry breaking mechanism may skip this subsection.
Let us focus on $SU(N)$ gauge theory defined in the five-dimensional space-time $M^4 \times (S^1/Z_2)$
where $M^4$ is the four-dimensional Minkowski space-time and $S^1/Z_2$ is the one-dimensional orbifold,
whose coordinates are denoted by $x^\mu$ (or $x$) and $y$, respectively.
The $S^1/Z_2$ is obtained by dividing the circle $S^1$ (with the identification $y\sim y+2\pi R$) 
by the $Z_2$ transformation $y \to -y$ so that the point $y$ is identified with $-y$.
Then the $S^1/Z_2$ is regarded as an interval with length $\pi R$, with $R$ being the $S^1$~radius.
The both end points $y = 0$ and $\pi R$ are fixed points under the $Z_2$ transformation.
For operations on the fifth coordinate defined by
\begin{align}
&\text{$Z_2$: $y \to -y$,} &
&\text{$Z'_2$: $y \to 2 \pi R - y$,} &
&\text{$T$: $y \to y + 2\pi R$,}
\label{Z2S}
\end{align}
the following relations hold:
\begin{align}
Z_2^2 &= {Z'_2}^2 = I , &
T     &= Z'_2 Z_2  ,
\label{Z2S-rel}
\end{align}
where $I$ is the identity operation.
The operation $Z'_2$ is the reflection at the end point $y = \pi R$
and the $S^1/Z_2$ can be defined as $R^1/(Z_2 \times Z'_2)$ by using operations $Z_2$ and $Z'_2$.

Although the point $y$ is identified with the points $-y$ and $2 \pi R - y$ on $S^1/Z_2$, 
a field does not necessarily take an identical value at these points.
We require that the Lagrangian density should be single-valued.
Then the following BCs of a field $\Phi(x,y)$ are allowed in general
\begin{align}
\Phi(x, -y)          &= T_{\Phi}[P_0] \Phi(x, y) , &
\Phi(x, 2 \pi R -y)  &= T_{\Phi}[P_1] \Phi(x, y) , &
\Phi(x, y + 2 \pi R) &= T_{\Phi}[U] \Phi(x, y) ,
\label{BC's-Phi}
\end{align}
where $T_{\Phi}[P_0]$, $T_{\Phi}[P_1]$ and $T_{\Phi}[U]$
represent appropriate representation matrices
including an arbitrary sign factors, with the matrices $P_0$, $P_1$ and $U$ 
standing for the representation matrices of the fundamental representation 
for the $Z_2$, $Z_2'$ and $T$ transformations (up to sign factors), respectively.
The representation matrices satisfy the counterparts of (\ref{Z2S-rel}):
\begin{align}
T_{\Phi}[P_0]^2 = T_{\Phi}[P_1]^2 &= \mathcal{I} , &
T_{\Phi}[U] &= T_{\Phi}[P_0] T_{\Phi}[P_1] ,
\label{Phi-rel}
\end{align}
where $\mathcal{I}$ stands for the unit matrix.
The eigenvalues of $T_{\Phi}[P_0]$ and $T_{\Phi}[P_1]$
are interpreted as the $Z_2$ parity for the fifth coordinate flip. 
As the assignment of $Z_2$ parity determines BCs of each multiplet on $S^1/Z_2$,
we use `$Z_2$ parity' as a parallel expression of `BCs on $S^1/Z_2$' in the remainder of the paper.

Let $\phi^{(\mathcal{P}_0 \mathcal{P}_1)}(x, y)$ be a component in a multiplet $\Phi(x, y)$ 
and have a definite $Z_2$ parity $(\mathcal{P}_0, \mathcal{P}_1)$. 
The Fourier-expansion of $\phi^{(\mathcal{P}_0 \mathcal{P}_1)}(x, y)$ is given by
\begin{eqnarray}
&~& \phi^{(++)}(x, y) = \frac{1}{\sqrt{\pi R}} \phi_0(x) + \sqrt{\frac{2}{\pi R}} \sum_{n=1}^{\infty} \phi_n(x) \cos \frac{ny}{R} ,
\label{phi++}\\
&~& \phi^{(+-)}(x, y) = \sqrt{\frac{2}{\pi R}} \sum_{n=1}^{\infty} \phi_n(x) \sin \frac{ny}{R} ,
\label{phi+-}\\
&~& \phi^{(-+)}(x, y) = \sqrt{\frac{2}{\pi R}} \sum_{n=1}^{\infty} \phi_n(x) \cos \frac{\left(n-\frac{1}{2}\right)y}{R} ,
\label{phi-+}\\
&~& \phi^{(--)}(x, y) = \sqrt{\frac{2}{\pi R}} \sum_{n=1}^{\infty} \phi_n(x) \sin \frac{\left(n-\frac{1}{2}\right)y}{R} ,
\label{phi--}
\end{eqnarray}
where $\pm$ indicates the eigenvalues $\pm1$ of $Z_2$ parity.
In the above Kaluza-Klein (KK) expansions~(\ref{phi++})--(\ref{phi--}), the coefficients~$\phi_m(x)$ ($m=0,1,\dots$)
are four-dimensional fields which acquire the KK mass~$m/R$ when the $Z_2$ parity is~$(+1, +1)$,
$n/R$ $(n=1,2,\dots)$ when~$(-1, -1)$, and $(n-\frac{1}{2})/R$ when~$(\pm1, \mp1)$ upon compactification.
{\it Unless all components of the non-singlet field have a common $Z_2$ parity, 
a symmetry reduction occurs upon compactification because zero modes $\phi_0(x)$ are absent in fields with an odd parity.}
This kind of symmetry breaking is called `orbifold breaking'.

Our four-dimensional world is assumed to be a boundary at one of the fixed points, on the basis of the `brane world scenario'.
There exist two kinds of four-dimensional field in our low-energy theory.
One is the brane field which lives only at the boundary and the other is the zero mode stemming from the bulk field.
The massive KK modes $\phi_n(x)$ do not appear in our low-energy world 
because they have heavy masses of $O(1/R)$, the magnitude same as the unification scale.
Chiral anomalies may arise at the boundaries with the advent of chiral fermions.
Those anomalies must be cancelled in the four-dimensional effective theory by the contribution of brane chiral fermions
and/or counter terms such as the Chern-Simons term~\cite{C&H,KK&L}.

\subsection{Orbifold breaking of $SU(N)$}

Now we prepare the basic building blocks for our argument.
For simplicity, we consider the symmetry breaking induced by the following representation matrices of the $Z_2$ parity
\begin{align}
P_0 &= \diag(\overbrace{+1, \dots, +1, +1, \dots, +1, -1, \dots, -1, -1, \dots, -1}^N) , 
\label{P0} \\
P_1 &= \diag(\underbrace{+1, \dots, +1}_{p}, \underbrace{-1, \dots, -1}_{q}, \underbrace{+1, \dots, +1}_{r}, 
 \underbrace{-1, \dots, -1}_{s}) ,
\label{P1}
\end{align}
where $s = N-p-q-r$.
The BCs (\ref{P0}) and (\ref{P1}) result in the symmetry breaking pattern $SU(N) \to  SU(p) \times SU(q) \times SU(r) \times SU(s) \times U(1)^\nu$.
Here and hereafter ``$SU(1)$'' unconventionally stands for $U(1)$, $SU(0)$ means nothing,
and $\nu = 3 - \kappa$ where $\kappa$ is the number of zero or one in $p$, $q$, $r$ and $s$.
The $Z_2$ parity (or BCs) specified by intergers $p$, $q$ and $r$ is also denoted $[p; q, r; s]$.

After the breakdown of~$SU(N)$, the rank~$k$ totally antisymmetric tensor representation~$[N, k]$, whose dimension is~${}_{N}C_{k}$,
is decomposed into a sum of multiplets of the subgroup $SU(p) \times SU(q) \times SU(r) \times SU(s)$ as
\begin{eqnarray}
[N, k] = \sum_{l_1 =0}^{k} \sum_{l_2 = 0}^{k-l_1} \sum_{l_3 = 0}^{k-l_1-l_2}  
\left({}_{p}C_{l_1}, {}_{q}C_{l_2}, {}_{r}C_{l_3}, {}_{s}C_{l_4}\right) ,
\label{Nk}
\end{eqnarray}
where $l_4=k-l_1-l_2-l_3$ and our notation is that ${}_{n}C_{l} = 0$ for $l > n$ and $l < 0$.
Here and hereafter we use ${}_{n}C_{l}$ instead of $[n, l]$ in many cases.
(We sometimes use the ordinary notation for representations too, 
e.g., ${\bf{5}}$ and ${\overline{\bf{5}}}$ in place of ${}_{5}C_{1}$ and ${}_{5}C_{4}$.) 

The $[N, k]$ is constructed by the antisymmetrization of $k$-ple product of the fundamental representation ${\bm{N}} = [N, 1]$:
\begin{eqnarray}
[N, k] = ({\bm{N}} \times \dots \times {\bm{N}})_a .
\label{N*...*N} 
\end{eqnarray}
We define the intrinsic $Z_2$ and $Z_2'$ parities $\eta_{[N,k]}$ and $\eta'_{[N,k]}$, respectively, such that
\begin{eqnarray}
&~& ({\bm{N}} \times \dots \times {\bm{N}})_a \to \eta_{[N,k]} (P_0 {\bm{N}} \times \dots \times P_0 {\bm{N}})_a ,~
\label{etaNk} \\
&~& ({\bm{N}} \times \dots \times {\bm{N}})_a \to \eta'_{[N,k]} (P_1 {\bm{N}} \times \dots \times P_1 {\bm{N}})_a .
\label{eta'Nk} 
\end{eqnarray}
By definition, $\eta_{[N,k]}$ and $\eta'_{[N,k]}$ take a value $+1$ or $-1$.
The $Z_2$ parity of the representation~$({}_{p}C_{l_1}, {}_{q}C_{l_2},$ ${}_{r}C_{l_3},$ ${}_{s}C_{l_4})$ is given by
\begin{eqnarray}
&~& \mathcal{P}_0 = (-1)^{l_3+l_4} \eta_{[N,k]} = (-1)^{l_1+l_2} (-1)^k \eta_{[N,k]} , 
\label{Z2}\\
&~& \mathcal{P}_1 = (-1)^{l_2+l_4} \eta'_{[N,k]} = (-1)^{l_1+l_3} (-1)^k \eta'_{[N,k]} .
\label{Z'2}
\end{eqnarray}

A fermion with spin $1/2$ in five dimensions is regarded as a Dirac fermion 
or a pair of Weyl fermions with opposite chiralities in four dimensions.
The representations of each Weyl fermions are decomposed as,
\begin{eqnarray}
&~& [N, k]_L = \sum_{l_1 =0}^{k} \sum_{l_2 = 0}^{k-l_1} \sum_{l_3 = 0}^{k-l_1-l_2}  
\left({}_{p}C_{l_1}, {}_{q}C_{l_2}, {}_{r}C_{l_3}, {}_{s}C_{l_4}\right)_L ,
\label{NkL}\\
&~& [N, k]_R = \sum_{l_1 =0}^{k} \sum_{l_2 = 0}^{k-l_1} \sum_{l_3 = 0}^{k-l_1-l_2}  
\left({}_{p}C_{l_1}, {}_{q}C_{l_2}, {}_{r}C_{l_3}, {}_{s}C_{l_4}\right)_R ,
\label{NkR}
\end{eqnarray}
where $l_4=k-l_1-l_2-l_3$.
Here the subscript $L$ ($R$) represents the left-handedness (right-handedness) for Weyl fermions.
The $Z_2$ parity of the representation~$({}_{p}C_{l_1}, {}_{q}C_{l_2},$ ${}_{r}C_{l_3},$ ${}_{s}C_{l_4})_L$ is given by
\begin{align}
\mathcal{P}_0 &= (-1)^{l_1+l_2} (-1)^k \eta_{[N,k]_L} , &
\mathcal{P}_1 &= (-1)^{l_1+l_3} (-1)^k \eta'_{[N,k]_L} .
\label{Z2L}
\end{align}
In the same way, the $Z_2$ parity of the representations~$({}_{p}C_{l_1}, {}_{q}C_{l_2},$ ${}_{r}C_{l_3},$ ${}_{s}C_{l_4})_R$ is given by
\begin{align}
\mathcal{P}_0 &= (-1)^{l_1+l_2} (-1)^k \eta_{[N,k]_R} , &
\mathcal{P}_1 &= (-1)^{l_1+l_3} (-1)^k \eta'_{[N,k]_R} .
\label{Z2R}
\end{align}
The $({}_{p}C_{l_1}, {}_{q}C_{l_2},$ ${}_{r}C_{l_3},$ ${}_{s}C_{l_4})_L$ and $({}_{p}C_{l_1}, {}_{q}C_{l_2},$ ${}_{r}C_{l_3},$ ${}_{s}C_{l_4})_R$ 
should have opposite $Z_2$ parity each other, $\eta_{[N,k]_R} = - \eta_{[N,k]_L}$ and  $\eta'_{[N,k]_R} = - \eta'_{[N,k]_L}$, 
from the requirement that the kinetic term is invariant under the $Z_2$ parity transformation.
The $Z_2$ transformation property for fermions is written down by
\begin{eqnarray}
&~& ({\bm{N}} \times \dots \times {\bm{N}})_a \to - \eta_{[N,k]_L} \gamma_5 (P_0 {\bm{N}} \times \dots \times P_0 {\bm{N}})_a ,~
\label{etaNkfermion} \\
&~& ({\bm{N}} \times \dots \times {\bm{N}})_a \to - \eta'_{[N,k]_L} \gamma_5 (P_1 {\bm{N}} \times \dots \times P_1 {\bm{N}})_a ,
\label{eta'Nkfermion} 
\end{eqnarray}
where $\gamma_5 \psi_L = -\psi_L$ and $\gamma_5 \psi_R = +\psi_R$. 
Hereafter we denote $\eta_{[N,k]_L}$ and $\eta'_{[N,k]_L}$ as $\eta_k$ and $\eta'_k$, respectively.
Not only left-handed Weyl fermions but also right-handed ones, having even $Z_2$ parities $\mathcal{P}_0 = \mathcal{P}_1 = +1$, compose chiral fermions in the SM.
At the same time, we want most zero modes of mirror particles to disappear.
In the next section, we will see that all zero modes of mirror particles can be eliminated by a good choice of $Z_2$ parity.

In SUSY models, the hypermultiplet is the fundamental quantity concerning bulk matter fields in five dimensions.
The hypermultiplet is equivalent to a pair of chiral multiplets 
with opposite gauge quantum numbers in four dimensions.
The chiral multiplet with the representation $[N,N-k]$, which is a conjugate of $[N,k]$, 
contains a left-handed Weyl fermion with $[N,N-k]_L$.
This Weyl fermion is regarded as a right-handed one with $[N,k]_R$ by the use of the charge conjugation.
Hence our analysis works on SUSY models as well as non-SUSY ones.

\section{Unification of quarks and leptons}

Towards grand unification of flavor, Georgi investigated decades ago whether no-repeated representations 
in $SU(N)$ models can lead to families based on the `survival hypothesis', 
and found that three families are derived from $[11, 4] + [11, 8] + [11, 9] + [11, 10]$ in $SU(11)$ model in four dimensions~\cite{GF}.
The survival hypothesis is the assumption that {\it if a symmetry is broken down to a smaller symmetry at a scale $M$,
then any fermion mass terms invariant under the smaller group induce fermion masses of order $O(M)$}~\cite{BNM&S}.
The analysis is quite interesting, but it has a limitation that an anomaly free set of representations must be chosen
as far as the space-time is assumed to be four-dimensional.
As we move from four dimensions to five dimensions, we are free from the limitation 
by assuming that the four-dimensional effective theory be anomaly free due to the presence of appropriate brane fields and/or
the Chern-Simons term.
It is meaningful to re-examine the idea of flavor unification using orbifold GUTs
and it is intriguing to answer the question whether (complete) family unification can be realized or not.

Let us investivate unification of quarks and leptons in two cases.
Each orbifold breaking pattern is given by
\begin{eqnarray}
&~& SU(N) \to  SU(5) \times SU(q) \times SU(r) \times SU(s) \times U(1)^\nu ,
\label{OB1}\\
&~& SU(N) \to  G_{SM} \times SU(r) \times SU(s) \times U(1)^{\nu-1} ,
\label{OB2}
\end{eqnarray}
where $G_{SM}$ is the SM gauge group, $SU(3)_C \times SU(2)_L \times U(1)_Y$.

\subsection{Family unification in $SU(N)\to SU(5)$ orbifold GUT}

We study the gauge symmetry breaking pattern 
$SU(N) \to  SU(5) \times SU(q) \times SU(r) \times SU(s) \times U(1)^\nu$,
realized by the following $Z_2$ parity assignment
\begin{align}
P_0 &= \diag(+1, +1, +1, +1, +1, +1, \dots, +1, -1, \dots, -1, -1, \dots, -1) , 
\label{P0-SU5} \\
P_1 &= \diag(+1, +1, +1, +1, +1, \underbrace{-1, \dots, -1}_{q}, 
\underbrace{+1, \dots, +1}_{r}, \underbrace{-1, \dots, -1}_{s}) ,
\label{P1-SU5}
\end{align}
where $s = N-5-q-r$.
After the breakdown of $SU(N)$, the totally antisymmetric representation $[N, k]$ is decomposed into
a sum of multiplets of the subgroup $SU(5) \times SU(q) \times SU(r) \times SU(s)$,
\begin{eqnarray}
[N, k] = \sum_{l_1 =0}^{k} \sum_{l_2 = 0}^{k-l_1} \sum_{l_3 = 0}^{k-l_1-l_2}  
\left({}_{5}C_{l_1}, {}_{q}C_{l_2}, {}_{r}C_{l_3}, {}_{s}C_{l_4}\right) ,
\label{Nk}
\end{eqnarray}
where $l_4=k-l_1-l_2-l_3$.
We list the $Z_2$ parity assignment for representations of Weyl fermions in Table \ref{t1}.
In the first column, species mean the representations of $SU(5)$.
As mentioned before, 
 ${{}_{5}C_{0}}$, ${{}_{5}C_{1}}$, ${{}_{5}C_{2}}$, ${{}_{5}C_{3}}$, ${{}_{5}C_{4}}$
and ${{}_{5}C_{5}}$ stand for representations ${\bf{1}}$, ${\bf{5}}$, ${\bf{10}}$, 
 ${\overline{\bf{10}}}$, ${\overline{\bf{5}}}$ and ${\overline{\bf{1}}}$.\footnote{
We denote the $SU(5)$ singlet relating to ${{}_{5}C_{5}}$ as ${\overline{\bf{1}}}$, for convenience sake,
to avoid the confusion over singlets.}

\begin{table}
\caption{The $Z_2$ parity assignment for representations of fermions.}
\label{t1}
\begin{center}
\begin{tabular}{c|c|c|c} \hline
{\it species} & {\it Representation}  & $\mathcal{P}_0$ & $\mathcal{P}_1$ \\ \hline\hline
${\bf 1}_L$ & $\left({}_{5}C_{0}, {}_{q}C_{l_2}, {}_{r}C_{l_3}, {}_{s}C_{k-l_2-l_3}\right)_L$  
 & $(-1)^{l_2} (-1)^k \eta_{k}$ & $(-1)^{l_3} (-1)^k \eta'_{k}$ \\ 
${\bf 1}_R$ & $\left({}_{5}C_{0}, {}_{q}C_{l_2}, {}_{r}C_{l_3}, {}_{s}C_{k-l_2-l_3}\right)_R$  
 & $-(-1)^{l_2} (-1)^k \eta_{k}$ & $-(-1)^{l_3} (-1)^k \eta'_{k}$ \\ \hline
${\bf 5}_L$ & $\left({}_{5}C_{1}, {}_{q}C_{l_2}, {}_{r}C_{l_3}, {}_{s}C_{k-l_2-l_3-1}\right)_L$  
 & $-(-1)^{l_2} (-1)^k \eta_{k}$ & $-(-1)^{l_3} (-1)^k \eta'_{k}$ \\ 
${\bf 5}_R$ & $\left({}_{5}C_{1}, {}_{q}C_{l_2}, {}_{r}C_{l_3}, {}_{s}C_{k-l_2-l_3-1}\right)_R$  
 & $(-1)^{l_2} (-1)^k \eta_{k}$ & $(-1)^{l_3} (-1)^k \eta'_{k}$ \\ \hline
${\bf 10}_L$ & $\left({}_{5}C_{2}, {}_{q}C_{l_2}, {}_{r}C_{l_3}, {}_{s}C_{k-l_2-l_3-2}\right)_L$  
 & $(-1)^{l_2} (-1)^k \eta_{k}$ & $(-1)^{l_3} (-1)^k \eta'_{k}$ \\ 
${\bf 10}_R$ & $\left({}_{5}C_{2}, {}_{q}C_{l_2}, {}_{r}C_{l_3}, {}_{s}C_{k-l_2-l_3-2}\right)_R$  
 & $-(-1)^{l_2} (-1)^k \eta_{k}$ & $-(-1)^{l_3} (-1)^k \eta'_{k}$ \\ \hline
${\overline{\bf 10}}_L$ & $\left({}_{5}C_{3}, {}_{q}C_{l_2}, {}_{r}C_{l_3}, {}_{s}C_{k-l_2-l_3-3}\right)_L$  
 & $-(-1)^{l_2} (-1)^k \eta_{k}$ & $-(-1)^{l_3} (-1)^k \eta'_{k}$ \\ 
${\overline{\bf 10}}_R$ & $\left({}_{5}C_{3}, {}_{q}C_{l_2}, {}_{r}C_{l_3}, {}_{s}C_{k-l_2-l_3-3}\right)_R$  
 & $(-1)^{l_2} (-1)^k \eta_{k}$ & $(-1)^{l_3} (-1)^k \eta'_{k}$ \\ \hline
${\overline{\bf 5}}_L$ & $\left({}_{5}C_{4}, {}_{q}C_{l_2}, {}_{r}C_{l_3}, {}_{s}C_{k-l_2-l_3-4}\right)_L$  
 & $(-1)^{l_2} (-1)^k \eta_{k}$ & $(-1)^{l_3} (-1)^k \eta'_{k}$ \\ 
${\overline{\bf 5}}_R$ & $\left({}_{5}C_{4}, {}_{q}C_{l_2}, {}_{r}C_{l_3}, {}_{s}C_{k-l_2-l_3-4}\right)_R$  
 & $-(-1)^{l_2} (-1)^k \eta_{k}$ & $-(-1)^{l_3} (-1)^k \eta'_{k}$ \\ \hline
${\overline{\bf 1}}_L$ & $\left({}_{5}C_{5}, {}_{q}C_{l_2}, {}_{r}C_{l_3}, {}_{s}C_{k-l_2-l_3}\right)_L$  
 & $-(-1)^{l_2} (-1)^k \eta_{k}$ & $-(-1)^{l_3} (-1)^k \eta'_{k}$ \\ 
${\overline{\bf 1}}_R$ & $\left({}_{5}C_{5}, {}_{q}C_{l_2}, {}_{r}C_{l_3}, {}_{s}C_{k-l_2-l_3}\right)_R$  
 & $(-1)^{l_2} (-1)^k \eta_{k}$ & $(-1)^{l_3} (-1)^k \eta'_{k}$ \\ \hline
\end{tabular}
\end{center}
\end{table}

Utilizing the survival hypothesis and the equivalence of $({\bf{5}}_R)^c$ and $(\overline{\bf{10}}_R)^c$
with $\overline{\bf{5}}_L$ and ${\bf{10}}_L$, respectively,%
\footnote{
	As usual, $({\bf{5}}_R)^c$ and $(\overline{\bf{10}}_R)^c$ represent 
the charge conjugate of ${\bf{5}}_R$ and $\overline{\bf{10}}_R$, respectively. 
Note that $({\bf{5}}_R)^c$ and $(\overline{\bf{10}}_R)^c$ transform as the left-handed Weyl fermions under the four dimensional Lorentz transformations.
	}
we write the numbers of $\overline{\bf 5}$ and ${\bf{10}}$ representations for left-handed Weyl fermions as
\begin{align}
n_{\bar{5}}
	&\equiv \sharp{\overline{\bf 5}}_L  - \sharp{\bf 5}_L 
  + \sharp{\bf 5}_R  - \sharp{\overline{\bf 5}}_R ,  
\label{nbar5-def}\\
n_{10} 
	&\equiv \sharp{\bf 10}_L  - \sharp{\overline{\bf 10}}_L 
  + \sharp{\overline{\bf 10}}_R  - \sharp{\bf 10}_R  , 
\label{n10-def}
\end{align}
where $\sharp$ represents the number of each multiplet.

When we take $\left((-1)^{k} \eta_{k}, (-1)^{k} \eta'_{k}\right)=\left(+1, +1\right)$, $n_{\bar{5}}$ and $n_{10}$ are given by
\begin{align}
n_{\bar{5}}
	&= \sum_{l_1 = 1, 4} \sum_{l_2 = 0, 2, \dots} \sum_{l_3 = 0, 2, \dots}
  {}_{q}C_{l_2} \cdot {}_{r}C_{l_3} \cdot {}_{s}C_{l_4} 
\nonumber \\
	&\quad\mbox{}- \sum_{l_1 = 1, 4} \sum_{l_2 = 1, 3, \dots} \sum_{l_3 = 1, 3, \dots}
  {}_{q}C_{l_2} \cdot {}_{r}C_{l_3} \cdot {}_{s}C_{l_4} \equiv n_{\bar{5},k}^{(++)} ,  
\label{nbar5++}\\
n_{10}
	&= \sum_{l_1 = 2, 3} \sum_{l_2 = 0, 2, \dots} \sum_{l_3 = 0, 2, \dots}
  {}_{q}C_{l_2} \cdot {}_{r}C_{l_3} \cdot {}_{s}C_{l_4} 
\nonumber \\
	&\quad\mbox{}- \sum_{l_1 = 2, 3} \sum_{l_2 = 1, 3, \dots} \sum_{l_3 = 1, 3, \dots}
  {}_{q}C_{l_2} \cdot {}_{r}C_{l_3} \cdot {}_{s}C_{l_4} \equiv n_{10,k}^{(++)},
\label{n10++}
\end{align}
where $s=N-5-q-r$ and $l_4=k-l_1-l_2-l_3$.
When we take $\left((-1)^{k} \eta_{k} , (-1)^{k} \eta'_{k}\right)=\left(+1, -1\right)$, $n_{\bar{5}}$ and $n_{10}$ are given by
\begin{align}
n_{\bar{5}}
	&= \sum_{l_1 = 1, 4} \sum_{l_2 = 0, 2, \dots} \sum_{l_3 = 1, 3, \dots}
  {}_{q}C_{l_2} \cdot {}_{r}C_{l_3} \cdot {}_{s}C_{l_4} 
\nonumber \\
	&\quad\mbox{}- \sum_{l_1 = 1, 4} \sum_{l_2 = 1, 3, \dots} \sum_{l_3 = 0, 2, \dots}
  {}_{q}C_{l_2} \cdot {}_{r}C_{l_3} \cdot {}_{s}C_{l_4} \equiv n_{\bar{5},k}^{(+-)} ,  
\label{nbar5+-}\\
n_{10}
	&= \sum_{l_1 = 2, 3} \sum_{l_2 = 0, 2, \dots} \sum_{l_3 = 1, 3, \dots}
  {}_{q}C_{l_2} \cdot {}_{r}C_{l_3} \cdot {}_{s}C_{l_4} 
\nonumber \\
	&\quad\mbox{}- \sum_{l_1 = 2, 3} \sum_{l_2 = 1, 3, \dots} \sum_{l_3 = 0, 2, \dots}
  {}_{q}C_{l_2} \cdot {}_{r}C_{l_3} \cdot {}_{s}C_{l_4} \equiv n_{10,k}^{(+-)},
\label{n10+-}
\end{align}
where $s$ and $l_4$ are defined same as above.
In the same way, we can derive $n_{\bar{5}} = -n_{\bar{5},k}^{(+-)}$ and $n_{10} = - n_{10,k}^{(+-)}$ 
for $\left((-1)^{k} \eta_{k}, (-1)^{k} \eta'_{k}\right)=\left(-1, +1\right)$
and $n_{\bar{5}} = -n_{\bar{5},k}^{(++)}$ and $n_{10} = - n_{10,k}^{(++)}$ 
for $\left((-1)^{k} \eta_{k}, (-1)^{k} \eta'_{k}\right)=\left(-1, -1\right)$.

First we consider the case with $q = 0$, i.e., $l_2 = 0$.
The intrinsic $Z_2$ parity should satisfy $\left((-1)^{k} \eta_{k} , (-1)^{k} \eta'_{k}\right)=\left(+1, +1\right)$ or $\left(+1, -1\right)$
from the requirement that $n_{\bar{5}}$ and $n_{10}$ are positive. 
{\it The ${\bf 5}_L$, ${\bf 10}_R$, ${\overline{\bf 10}}_L$ and ${\overline{\bf 5}}_R$ are regarded as mirror particles
and all those zero modes are projected out by the assignment $(-1)^{k} \eta_{k} = +1$.}
Then $n_{\bar{5}}$ and $n_{10}$ are simplified to 
\begin{eqnarray}
&~& n_{\bar{5},k}^{(++)}(q=0) = \sum_{l_1 = 1, 4} \sum_{l_3 = 0, 2, \dots} {}_{r}C_{l_3} \cdot {}_{N-5-r}C_{k-l_1-l_3} ,  
\label{nbar5++q=0}\\
&~& n_{10,k}^{(++)}(q=0) = \sum_{l_1 = 2, 3} \sum_{l_3 = 0, 2, \dots} {}_{r}C_{l_3} \cdot {}_{N-5-r}C_{k-l_1-l_3} 
\label{n10++q=0}
\end{eqnarray}
and 
\begin{eqnarray}
&~& n_{\bar{5},k}^{(+-)}(q=0) = \sum_{l_1 = 1, 4} \sum_{l_3 = 1, 3, \dots} {}_{r}C_{l_3} \cdot {}_{N-5-r}C_{k-l_1-l_3} ,  
\label{nbar5+-q=0}\\
&~& n_{10,k}^{(+-)}(q=0) = \sum_{l_1 = 2, 3} \sum_{l_3 = 1, 3, \dots} {}_{r}C_{l_3} \cdot {}_{N-5-r}C_{k-l_1-l_3} .
\label{n10+-q=0}
\end{eqnarray}
Three families appear from $[9, 6]$ when $r=3$, using
\begin{eqnarray}
\hspace{-1cm} &~& n_{\bar{5},k}^{(++)}(q=0, r=3) = {}_{N-8}C_{k-1} + 3{}_{N-8}C_{k-3} + {}_{N-8}C_{k-4} + 3{}_{N-8}C_{k-6} ,  
\label{nbar5++q=0r=3}\\
\hspace{-1cm} &~& n_{10,k}^{(++)}(q=0, r=3) = {}_{N-8}C_{k-2} + 3{}_{N-8}C_{k-4} + {}_{N-8}C_{k-3} + 3{}_{N-8}C_{k-5} 
\label{n10++q=0r=3}
\end{eqnarray}
and from $[9, 3]$ when $r=3$, using
\begin{eqnarray}
\hspace{-1cm} &~& n_{\bar{5},k}^{(+-)}(q=0, r=3) = 3{}_{N-8}C_{k-2} + {}_{N-8}C_{k-4} + 3{}_{N-8}C_{k-5} + {}_{N-8}C_{k-7} ,  
\label{nbar5+-q=0r=3}\\
\hspace{-1cm} &~& n_{10,k}^{(+-)}(q=0, r=3) = 3{}_{N-8}C_{k-3} + {}_{N-8}C_{k-5} + 3{}_{N-8}C_{k-4} + {}_{N-8}C_{k-6} .
\label{n10+-q=0r=3}
\end{eqnarray}
We find that there is no other solutions with $q=0$ in which three families $n_{\bar{5}} = n_{10} = 3$ are originating from a single representation to achieve the complete family unification.

There are many other possibilities to derive three families when we consider the case with $q \ne 0$.
We list representations and BCs to derive three families $n_{\bar{5}} = n_{10} = 3$ up to $SU(15)$ in Table \ref{t2}.
\begin{table}
\vspace{-0.6cm}
\caption{Representations and BCs derived three families up to $SU(15)$.}
\label{t2}
\begin{center}
\begin{tabular}{c|c|c|c} \hline
 Representation & $[p; q, r; s]$ & $(-1)^{k} \eta_{k}$ & $(-1)^k \eta'_{k}$ \\ \hline\hline
 $[9,3]$ & $[5; 0, 3; 1]$ & $+1$ & $-1$ \\ 
 $[9,3]$ & $[5; 3, 0; 1]$ & $-1$ & $+1$ \\ 
 $[9,6]$ & $[5; 0, 3; 1]$ & $+1$ & $+1$ \\ 
 $[9,6]$ & $[5; 3, 0; 1]$ & $+1$ & $+1$ \\ \hline
 $[11,3]$ & $[5; 1, 4; 1]$ & $+1$ & $-1$ \\ 
 $[11,3]$ & $[5; 4, 1; 1]$ & $-1$ & $+1$ \\ 
 $[11,4]$ & $[5; 1, 4; 1]$ & $+1$ & $+1$ \\ 
 $[11,4]$ & $[5; 4, 1; 1]$ & $+1$ & $+1$ \\ 
 $[11,7]$ & $[5; 1, 4; 1]$ & $-1$ & $+1$ \\ 
 $[11,7]$ & $[5; 4, 1; 1]$ & $+1$ & $-1$ \\ 
 $[11,8]$ & $[5; 1, 4; 1]$ & $-1$ & $-1$ \\ 
 $[11,8]$ & $[5; 4, 1; 1]$ & $-1$ & $-1$ \\ \hline
 $[12,3]$ & $[5; 1, 4; 2]$ & $+1$ & $+1$ \\ 
 $[12,3]$ & $[5; 4, 1; 2]$ & $+1$ & $+1$ \\ 
 $[12,9]$ & $[5; 1, 4; 2]$ & $-1$ & $+1$ \\ 
 $[12,9]$ & $[5; 4, 1; 2]$ & $+1$ & $-1$ \\ \hline
 $[13,3]$ & $[5; 2, 5; 1]$ & $+1$ & $-1$ \\ 
 $[13,3]$ & $[5; 5, 2; 1]$ & $-1$ & $+1$ \\ 
 $[13,10]$ & $[5; 2, 5; 1]$ & $+1$ & $+1$ \\ 
 $[13,10]$ & $[5; 5, 2; 1]$ & $+1$ & $+1$ \\ \hline
 $[14,4]$ & $[5; 4, 4; 1]$ & $-1$ & $-1$ \\ 
 $[14,10]$ & $[5; 2, 6; 1]$ & $+1$ & $+1$ \\ 
 $[14,10]$ & $[5; 4, 4; 1]$ & $-1$ & $-1$ \\ 
 $[14,10]$ & $[5; 6, 2; 1]$ & $+1$ & $+1$ \\ \hline
 $[15,3]$ & $[5; 3, 6; 1]$ & $+1$ & $-1$ \\ 
 $[15,3]$ & $[5; 6, 3; 1]$ & $-1$ & $+1$ \\ 
 $[15,4]$ & $[5; 4, 5; 1]$ & $-1$ & $-1$ \\ 
 $[15,4]$ & $[5; 5, 4; 1]$ & $-1$ & $-1$ \\ 
 $[15,5]$ & $[5; 4, 5; 1]$ & $-1$ & $+1$ \\ 
 $[15,5]$ & $[5; 5, 4; 1]$ & $+1$ & $-1$ \\ 
 $[15,10]$ & $[5; 4, 5; 1]$ & $-1$ & $-1$ \\ 
 $[15,10]$ & $[5; 5, 4; 1]$ & $-1$ & $-1$ \\ 
 $[15,11]$ & $[5; 3, 6; 1]$ & $+1$ & $-1$ \\ 
 $[15,11]$ & $[5; 4, 5; 1]$ & $-1$ & $+1$ \\ 
 $[15,11]$ & $[5; 5, 4; 1]$ & $+1$ & $-1$ \\ 
 $[15,11]$ & $[5; 6, 3; 1]$ & $-1$ & $+1$ \\ \hline
\end{tabular}
\end{center}
\end{table}
Further three family models exist if we allow several representations and/or brane fields.
For example, three families are generated from $[7,1] + [7,2] + [7,3] + [7,4]$ or $[8,5] + [8,6]$ 
for $q = s = 0$ and $\left((-1)^{k} \eta_{k} , (-1)^{k} \eta'_{k}\right)=\left(+1, +1\right)$.

The $SU(5)$ singlets are regarded as the so-called right-handed neutrinos 
which can obtain heavy Majorana masses among themselves as well as the Dirac masses with left-handed neutrinos.
Some of them can be involved in see-saw mechanism~\cite{see-saw}.
From the definition of the total number of $SU(5)$ singlets (with heavy masses)
\begin{eqnarray}
&~& n_{1} \equiv \sharp  {}_{5}C_{0L}  + \sharp  {}_{5}C_{5L} 
  + \sharp  {}_{5}C_{5R} + \sharp  {}_{5}C_{0R}  , 
\label{n1-def}
\end{eqnarray}
the numbers for our cases are given by
\begin{align}
n_{1,k}^{(++)}
	&= \sum_{l_1 = 0, 5} \sum_{l_2 = 0, 2, \dots} \sum_{l_3 = 0, 2, \dots}
  {}_{q}C_{l_2} \cdot {}_{r}C_{l_3} \cdot {}_{s}C_{l_4} 
\nonumber \\
	&\quad\mbox{}+ \sum_{l_1 = 0, 5} \sum_{l_2 = 1, 3, \dots} \sum_{l_3 = 1, 3, \dots}
  {}_{q}C_{l_2} \cdot {}_{r}C_{l_3} \cdot {}_{s}C_{l_4}
\label{n1++}\\
n_{1,k}^{(+-)} 
	&= \sum_{l_1 = 0, 5} \sum_{l_2 = 0, 2, \dots} \sum_{l_3 = 1, 3, \dots}
  {}_{q}C_{l_2} \cdot {}_{r}C_{l_3} \cdot {}_{s}C_{l_4} 
\nonumber \\
	&\quad\mbox{}+ \sum_{l_1 = 0, 5} \sum_{l_2 = 1, 3, \dots} \sum_{l_3 = 0, 2, \dots}
  {}_{q}C_{l_2} \cdot {}_{r}C_{l_3} \cdot {}_{s}C_{l_4},
\label{n1+-}
\end{align}
where $s=N-5-q-r$ and $l_4=k-l_1-l_2-l_3$, for the corresponding intrinsic parity assignments.
Using Eqs.~(\ref{n1++}) and (\ref{n1+-}), we obtain one $SU(5)$ singlet
from $[9,6]$ when $\left(\eta_{6}, \eta'_{6}\right)=\left(+1, +1\right)$  
and from $[9,3]$ when $\left(-\eta_{3}, -\eta'_{3}\right)=\left(+1, -1\right)$ in the case that $q = 0$ and $r = 3$.
Other neutrino singlets might be supplied as brane fields.

We have studied the case with $p = 5$. 
The other cases with $q=5$, $r=5$ or $s=5$ are equivalent to that with $p=5$.
For example, the choice with $q = 5$ is equivalent to that with $p = 5$ by the exchange of
variables $p \leftrightarrow q$, $r \leftrightarrow s$ 
and the sign change of the intrinsic $Z_2$ parity $\eta'_{k} \leftrightarrow -\eta'_{k}$.
In the same way, the choice with $r = 5$ is equivalent to that with $p = 5$ by the exchange of
variables $p \leftrightarrow r$, $q \leftrightarrow s$ and the sign change $\eta_{k} \leftrightarrow -\eta_{k}$.
The choice with $s = 5$ is equivalent to that with $p = 5$ by the exchange of
variables $p \leftrightarrow s$, $q \leftrightarrow r$ 
and the sign change $\eta_{k} \leftrightarrow -\eta_{k}$ and $\eta'_{k} \leftrightarrow -\eta'_{k}$.

\subsection{Family unification in $SU(N)$ orbifold GUT directly broken to SM}

Next we study the gauge symmetry breaking pattern, 
$SU(N) \to  SU(3) \times SU(2) \times SU(r) \times SU(s) \times U(1)^\nu$,
which is realized by the $Z_2$ parity assignment
\begin{align}
P_0 &= \diag(+1, +1, +1, +1, +1, -1, \dots, -1, -1, \dots, -1) ~, 
\label{P0-GSM} \\
P_1 &= \diag(+1, +1, +1, -1, -1, \underbrace{+1, \dots, +1}_{r}, \underbrace{-1, \dots, -1}_{s}) ,
\label{P1-GSM}
\end{align}
where $s = N-5-r$ and $N \ge 6$.
After the breakdown of $SU(N)$, the totally antisymmetric representation $[N, k]$ are decomposed into
a sum of multiplets of the subgroup $SU(3) \times SU(2) \times SU(r) \times SU(s)$
\begin{eqnarray}
[N, k] = \sum_{l_1 =0}^{k} \sum_{l_2 = 0}^{k-l_1} \sum_{l_3 = 0}^{k-l_1-l_2}  
\left({}_{3}C_{l_1}, {}_{2}C_{l_2}, {}_{r}C_{l_3}, {}_{s}C_{l_4}\right) ,
\label{Nk}
\end{eqnarray}
where $l_1$, $l_2$ and $l_3$ are intergers and $l_4=k-l_1-l_2-l_3$.
We list $U(1)$ charges for representations in Table \ref{t3}.
The $U(1)$ charges are those in the following subgroups,
\begin{align}
SU(5)
	\supset\, & SU(3) \times SU(2) \times U(1)_1 , \\
SU(N-5)
	\supset\, & SU(r) \times SU(N-5-r) \times U(1)_2 , \nonumber \\
	        & SU(N-5-1) \times U(1)_2 ,  \\
SU(N)
	\supset\, & SU(5) \times SU(N-5) \times U(1)_3,
\end{align}
up to normalization.
We assume that $G_{SM} = SU(3) \times SU(2) \times U(1)_1$ up to normalization of the hypercharge.
Particle species are identified with the SM fermions by the gauge quantum numbers.
The particles with prime are regarded as mirror particles and expected to have no zero modes.
Each fermion has a definite chirality, e.g. $(d_{R})^c$ is left-handed and $d_{R}$ is right-handed.
\begin{table}
\caption{The $U(1)$ charges for representations of fermions.}
\label{t3}
\begin{center}
\begin{tabular}{c|c|c|c|c} \hline
{\it species} & {\it Representation}  & $U(1)_1$ & $U(1)_2$ & $U(1)_3$ \\ \hline\hline
$(\nu_{R})^c$, $\hat{\nu}_{R}$ & $\left({}_{3}C_{0}, {}_{2}C_{0}, {}_{r}C_{l_3}, {}_{s}C_{k-l_3}\right)$ & $0$ & $(N-5)l_3 - rk$ & $-5k$ \\ \hline
$(d'_{R})^c$, $d_{R}$ & $\left({}_{3}C_{1}, {}_{2}C_{0}, {}_{r}C_{l_3}, {}_{s}C_{k-l_3-1}\right)$ & $-2$ & $(N-5)l_3 - r(k-1)$ & $N-5k$ \\
$l'_{L}$, $(l_{L})^c$ & $\left({}_{3}C_{0}, {}_{2}C_{1}, {}_{r}C_{l_3}, {}_{s}C_{k-l_3-1}\right)$ & $3$ & $(N-5)l_3 - r(k-1)$ & $N-5k$ \\ \hline
$(u_{R})^c$, $u'_{R}$ & $\left({}_{3}C_{2}, {}_{2}C_{0}, {}_{r}C_{l_3}, {}_{s}C_{k-l_3-2}\right)$ & $-4$ & $(N-5)l_3 - r(k-2)$ & $2N-5k$ \\
$(e_{R})^c$, $e'_{R}$ & $\left({}_{3}C_{0}, {}_{2}C_{2}, {}_{r}C_{l_3}, {}_{s}C_{k-l_3-2}\right)$ & $6$ & $(N-5)l_3 - r(k-2)$ & $2N-5k$ \\
$q_{L}$, $(q'_{L})^c$ & $\left({}_{3}C_{1}, {}_{2}C_{1}, {}_{r}C_{l_3}, {}_{s}C_{k-l_3-2}\right)$ & $1$ & $(N-5)l_3 - r(k-2)$ & $2N-5k$ \\ \hline
$(e'_{R})^c$, $e_{R}$ & $\left({}_{3}C_{3}, {}_{2}C_{0}, {}_{r}C_{l_3}, {}_{s}C_{k-l_3-3}\right)$ & $-6$ & $(N-5)l_3 - r(k-3)$ & $3N-5k$ \\
$(u'_{R})^c$, $u_{R}$ & $\left({}_{3}C_{1}, {}_{2}C_{2}, {}_{r}C_{l_3}, {}_{s}C_{k-l_3-3}\right)$ & $4$ & $(N-5)l_3 - r(k-3)$ & $3N-5k$ \\
$q'_{L}$, $(q_{L})^c$ & $\left({}_{3}C_{2}, {}_{2}C_{1}, {}_{r}C_{l_3}, {}_{s}C_{k-l_3-3}\right)$ & $-1$ & $(N-5)l_3 - r(k-3)$ & $3N-5k$ \\ \hline
$l_{L}$, $(l'_{L})^c$ & $\left({}_{3}C_{3}, {}_{2}C_{1}, {}_{r}C_{l_3}, {}_{s}C_{k-l_3-4}\right)$ & $-3$ & $(N-5)l_3 - r(k-4)$ & $4N-5k$ \\
$(d_{R})^c$, $d'_{R}$ & $\left({}_{3}C_{2}, {}_{2}C_{2}, {}_{r}C_{l_3}, {}_{s}C_{k-l_3-4}\right)$ & $2$ & $(N-5)l_3 - r(k-4)$ & $4N-5k$ \\ \hline
$(\hat{\nu}_{R})^c$, $\nu_{R}$ & $\left({}_{3}C_{3}, {}_{2}C_{2}, {}_{r}C_{l_3}, {}_{s}C_{k-l_3-5}\right)$ & $0$ & $(N-5)l_3 - r(k-5)$ & $5N-5k$ \\ \hline
\end{tabular}
\end{center}
\end{table}
We list the $Z_2$ parity assignment for species in Table \ref{t4}.
Note that mirror particles have the $Z_2$ parity $\mathcal{P}_0 = -(-1)^k \eta_{k}$.

\begin{table}
\caption{The $Z_2$ parity assignment for representations of fermions.}
\label{t4}
\begin{center}
\begin{tabular}{c|c|c|c} \hline
{\it species} & {\it Representation}  & $\mathcal{P}_0$ & $\mathcal{P}_1$ \\ \hline\hline
$(\nu_{R})^c$ & $\left({}_{3}C_{0}, {}_{2}C_{0}, {}_{r}C_{l_3}, {}_{s}C_{k-l_3}\right)_L$  
 & $(-1)^k \eta_{k}$ & $(-1)^{l_3} (-1)^k \eta'_{k}$ \\ 
$\hat{\nu}_{R}$ & $\left({}_{3}C_{0}, {}_{2}C_{0}, {}_{r}C_{l_3}, {}_{s}C_{k-l_3}\right)_R$  
 & $-(-1)^k \eta_{k}$ & $-(-1)^{l_3} (-1)^k \eta'_{k}$ \\ \hline
$(d'_{R})^c$ & $\left({}_{3}C_{1}, {}_{2}C_{0}, {}_{r}C_{l_3}, {}_{s}C_{k-l_3-1}\right)_L$  
 & $-(-1)^k \eta_{k}$ & $-(-1)^{l_3} (-1)^k \eta'_{k}$ \\
$l'_{L}$ & $\left({}_{3}C_{0}, {}_{2}C_{1}, {}_{r}C_{l_3}, {}_{s}C_{k-l_3-1}\right)_L$  
 & $-(-1)^k \eta_{k}$  & $(-1)^{l_3} (-1)^k \eta'_{k}$ \\ 
$d_{R}$ & $\left({}_{3}C_{1}, {}_{2}C_{0}, {}_{r}C_{l_3}, {}_{s}C_{k-l_3-1}\right)_R$  
 & $(-1)^k \eta_{k}$ & $(-1)^{l_3} (-1)^k \eta'_{k}$ \\
$(l_{L})^c$ & $\left({}_{3}C_{0}, {}_{2}C_{1}, {}_{r}C_{l_3}, {}_{s}C_{k-l_3-1}\right)_R$  
 & $(-1)^k \eta_{k}$  & $-(-1)^{l_3} (-1)^k \eta'_{k}$ \\ \hline
$(u_{R})^c$ & $\left({}_{3}C_{2}, {}_{2}C_{0}, {}_{r}C_{l_3}, {}_{s}C_{k-l_3-2}\right)_L$ 
 & $(-1)^k \eta_{k}$  & $(-1)^{l_3} (-1)^k \eta'_{k}$ \\
$(e_{R})^c$ & $\left({}_{3}C_{0}, {}_{2}C_{2}, {}_{r}C_{l_3}, {}_{s}C_{k-l_3-2}\right)_L$ 
 & $(-1)^k \eta_{k}$  & $(-1)^{l_3} (-1)^k \eta'_{k}$ \\
$q_{L}$ & $\left({}_{3}C_{1}, {}_{2}C_{1}, {}_{r}C_{l_3}, {}_{s}C_{k-l_3-2}\right)_L$ 
 & $(-1)^k \eta_{k}$  & $-(-1)^{l_3} (-1)^k \eta'_{k}$ \\ 
$u'_{R}$ & $\left({}_{3}C_{2}, {}_{2}C_{0}, {}_{r}C_{l_3}, {}_{s}C_{k-l_3-2}\right)_R$ 
 & $-(-1)^k \eta_{k}$  & $-(-1)^{l_3} (-1)^k \eta'_{k}$ \\
$e'_{R}$ & $\left({}_{3}C_{0}, {}_{2}C_{2}, {}_{r}C_{l_3}, {}_{s}C_{k-l_3-2}\right)_R$ 
 & $-(-1)^k \eta_{k}$  & $-(-1)^{l_3} (-1)^k \eta'_{k}$ \\
$(q'_{L})^c$ & $\left({}_{3}C_{1}, {}_{2}C_{1}, {}_{r}C_{l_3}, {}_{s}C_{k-l_3-2}\right)_R$ 
 & $-(-1)^k \eta_{k}$  & $(-1)^{l_3} (-1)^k \eta'_{k}$ \\ \hline
$(e'_{R})^c$ & $\left({}_{3}C_{3}, {}_{2}C_{0}, {}_{r}C_{l_3}, {}_{s}C_{k-l_3-3}\right)_L$ 
 & $-(-1)^k \eta_{k}$  & $-(-1)^{l_3} (-1)^k \eta'_{k}$ \\
$(u'_{R})^c$ & $\left({}_{3}C_{1}, {}_{2}C_{2}, {}_{r}C_{l_3}, {}_{s}C_{k-l_3-3}\right)_L$ 
 & $-(-1)^k \eta_{k}$  & $-(-1)^{l_3} (-1)^k \eta'_{k}$ \\
$q'_{L}$ & $\left({}_{3}C_{2}, {}_{2}C_{1}, {}_{r}C_{l_3}, {}_{s}C_{k-l_3-3}\right)_L$ 
 & $-(-1)^k \eta_{k}$  & $(-1)^{l_3} (-1)^k \eta'_{k}$ \\ 
$e_{R}$ & $\left({}_{3}C_{3}, {}_{2}C_{0}, {}_{r}C_{l_3}, {}_{s}C_{k-l_3-3}\right)_R$ 
 & $(-1)^k \eta_{k}$  & $(-1)^{l_3} (-1)^k \eta'_{k}$ \\
$u_{R}$ & $\left({}_{3}C_{1}, {}_{2}C_{2}, {}_{r}C_{l_3}, {}_{s}C_{k-l_3-3}\right)_R$ 
 & $(-1)^k \eta_{k}$  & $(-1)^{l_3} (-1)^k \eta'_{k}$ \\
$(q_{L})^c$ & $\left({}_{3}C_{2}, {}_{2}C_{1}, {}_{r}C_{l_3}, {}_{s}C_{k-l_3-3}\right)_R$ 
 & $(-1)^k \eta_{k}$  & $-(-1)^{l_3} (-1)^k \eta'_{k}$ \\ \hline
$l_{L}$ & $\left({}_{3}C_{3}, {}_{2}C_{1}, {}_{r}C_{l_3}, {}_{s}C_{k-l_3-4}\right)_L$ 
 & $(-1)^k \eta_{k}$  & $-(-1)^{l_3} (-1)^k \eta'_{k}$ \\
$(d_{R})^c$ & $\left({}_{3}C_{2}, {}_{2}C_{2}, {}_{r}C_{l_3}, {}_{s}C_{k-l_3-4}\right)_L$ 
 & $(-1)^k \eta_{k}$  & $(-1)^{l_3} (-1)^k \eta'_{k}$ \\ 
$(l'_{L})^c$ & $\left({}_{3}C_{3}, {}_{2}C_{1}, {}_{r}C_{l_3}, {}_{s}C_{k-l_3-4}\right)_R$ 
 & $-(-1)^k \eta_{k}$  & $(-1)^{l_3} (-1)^k \eta'_{k}$ \\
$d'_{R}$ & $\left({}_{3}C_{2}, {}_{2}C_{2}, {}_{r}C_{l_3}, {}_{s}C_{k-l_3-4}\right)_R$ 
 & $-(-1)^k \eta_{k}$  & $-(-1)^{l_3} (-1)^k \eta'_{k}$ \\ \hline
$(\hat{\nu}_{R})^c$ & $\left({}_{3}C_{3}, {}_{2}C_{2}, {}_{r}C_{l_3}, {}_{s}C_{k-l_3-5}\right)_L$ 
 & $-(-1)^k \eta_{k}$  & $-(-1)^{l_3} (-1)^k \eta'_{k}$ \\ 
$\nu_{R}$ & $\left({}_{3}C_{3}, {}_{2}C_{2}, {}_{r}C_{l_3}, {}_{s}C_{k-l_3-5}\right)_R$ 
 & $(-1)^k \eta_{k}$  & $(-1)^{l_3} (-1)^k \eta'_{k}$ \\ \hline
\end{tabular}
\end{center}
\end{table}

The flavor numbers of down-type anti-quark singlets $(d_{R})^c$, lepton doublets $l_{L}$, up-type anti-quark singlets $(u_{R})^c$, 
positron-type lepton singlets $(e_{R})^c$, 
and quark doublets $q_{L}$ are denoted as $n_{\bar{d}}$, $n_l$, $n_{\bar{u}}$, $n_{\bar{e}}$ and $n_q$.
Using the survival hypothesis and the equivalence on charge conjugation, 
we define the flavor number of each chiral fermion as
\begin{align}
n_{\bar{d}} 
	&\equiv \sharp  ({}_{3}C_{2}, {}_{2}C_{2})_L - \sharp  ({}_{3}C_{1}, {}_{2}C_{0})_L 
	+ \sharp  ({}_{3}C_{1}, {}_{2}C_{0})_R - \sharp  ({}_{3}C_{2}, {}_{2}C_{2})_R ,  
\label{nd-def}\\
n_{l} 
	&\equiv \sharp  ({}_{3}C_{3}, {}_{2}C_{1})_L  - \sharp  ({}_{3}C_{0}, {}_{2}C_{1})_L 
	+ \sharp  ({}_{3}C_{0}, {}_{2}C_{1})_R  - \sharp  ({}_{3}C_{3}, {}_{2}C_{1})_R ,  
\label{nl-def}\\
n_{\bar{u}} 
	&\equiv \sharp  ({}_{3}C_{2}, {}_{2}C_{0})_L  - \sharp  ({}_{3}C_{1}, {}_{2}C_{2})_L 
	+ \sharp  ({}_{3}C_{1}, {}_{2}C_{2})_R  - \sharp  ({}_{3}C_{2}, {}_{2}C_{0})_R ,  
\label{nu-def}\\
n_{\bar{e}} 
	&\equiv \sharp  ({}_{3}C_{0}, {}_{2}C_{2})_L  - \sharp  ({}_{3}C_{3}, {}_{2}C_{0})_L 
	+ \sharp  ({}_{3}C_{3}, {}_{2}C_{0})_R  - \sharp  ({}_{3}C_{0}, {}_{2}C_{2})_R ,  
\label{ne-def}\\
n_{q} 
	&\equiv \sharp  ({}_{3}C_{1}, {}_{2}C_{1})_L  - \sharp  ({}_{3}C_{2}, {}_{2}C_{1})_L 
	+ \sharp  ({}_{3}C_{2}, {}_{2}C_{1})_R  - \sharp  ({}_{3}C_{1}, {}_{2}C_{1})_R ,  
\label{nq-def}
\end{align}
where $\sharp$ again represents the number of each multiplet.
{\it When we take $(-1)^k \eta_{k} = +1$, all zero modes of mirror particles are projected out.}
Hereafter we consider such a case.
The total number of (heavy) neutrino singlets $(\nu_{R})^c$ is denoted $n_{\bar{\nu}}$ and defined as
\begin{align}
n_{\bar{\nu}}
	&\equiv \sharp  ({}_{3}C_{0}, {}_{2}C_{0})_L  + \sharp  ({}_{3}C_{3}, {}_{2}C_{2})_L 
	+ \sharp  ({}_{3}C_{3}, {}_{2}C_{2})_R  + \sharp  ({}_{3}C_{0}, {}_{2}C_{0})_R .  
\label{nnu-def}
\end{align}

When we take $(-1)^k \eta'_{k} = +1$,  the numbers (\ref{nd-def})--(\ref{nnu-def}) are simplified to 
\begin{align}
n_{\bar{d}} 
	&= \sum_{i = 1, 4} \sum_{l_3 = 0, 2, \dots} {}_{r}C_{l_3} \cdot {}_{N-5-r}C_{k-i-l_3} \equiv n_{\bar{d},k}^{(++)},
\label{nd}\\
n_{l}
	&= \sum_{i = 1, 4} \sum_{l_3 = 1, 3, \dots} {}_{r}C_{l_3} \cdot {}_{N-5-r}C_{k-i-l_3} \equiv n_{l,k}^{(++)},
\label{nl}\\
n_{\bar{u}} = n_{\bar{e}} 
	&= \sum_{i = 2, 3} \sum_{l_3 = 0, 2, \dots} {}_{r}C_{l_3} \cdot {}_{N-5-r}C_{k-i-l_3} 
\equiv n_{\bar{e},k}^{(++)}, n_{\bar{u},k}^{(++)} ,
\label{ne}\\
n_{q} 
	&= \sum_{i = 2, 3} \sum_{l_3 = 1, 3, \dots} {}_{r}C_{l_3} \cdot {}_{N-5-r}C_{k-i-l_3} \equiv n_{q,k}^{(++)} ,
\label{nq}\\
n_{\bar{\nu}} 
	&= \sum_{i = 0, 5} \sum_{l_3 = 0, 2, \dots} {}_{r}C_{l_3} \cdot {}_{N-5-r}C_{k-i-l_3} 
\equiv n_{\bar{\nu},k}^{(++)} .
\label{nnu}
\end{align}
By comparing with the numbers (\ref{nbar5++q=0})--(\ref{n10+-q=0}) and (\ref{n1+-}), we obtain the following relations,
\begin{align}
n_{\bar{d},k}^{(++)} 
	&= n_{\bar{5},k}^{(++)}(q=0) , &
n_{l,k}^{(++)} 
	&= n_{\bar{5},k}^{(+-)}(q=0) , &
\label{nld-5}\\
n_{\bar{u},k}^{(++)} = n_{\bar{e},k}^{(++)} 
	&= n_{10,k}^{(++)}(q=0) ,      &
n_{q,k}^{(++)}
	&= n_{10,k}^{(+-)}(q=0) ,
\label{neuq-10}\\
n_{\bar{\nu},k}^{(++)} 
	&= n_{1,k}^{(++)}(q=0) .
\label{nnu-1}
\end{align}
When we take $(-1)^k \eta'_{k} = -1$, we obtain formulae in which $n_l$ is exchanged by $n_{\bar{d}}$
and $n_q$ by $n_{\bar{u}}$ ($n_{\bar{e}}$) in Eqs.~(\ref{nl})--(\ref{nq}).
The total number of (heavy) neutrino singlets is given by
$n_{\bar{\nu},k}^{(+-)} = n_{1k}^{(+-)}(q=0) = \sum_{i = 0, 5} \sum_{l_3 = 1, 3, \dots} {}_{r}C_{l_3} \cdot {}_{N-5-r}C_{k-i-l_3}$.
{}From the result in the previous subsection, we find that {\it there is no solution satisfying
$n_{\bar{d}} = n_{l} = n_{\bar{u}} = n_{\bar{e}} = n_{q} = n_{\bar{\nu}} = 3$.} 
We list the flavor number of each chiral fermion derived from the representation $[8, k]$ and $[9, k]$ 
for $r = 3$ in Table \ref{t5} and \ref{t6}, respectively.

\begin{table}
\caption{The flavor number of each chiral fermion from $[8, k]$ for $r = 3$.}
\label{t5}
\begin{center}
\begin{tabular}{c|c|c|c|c|c|c|c|c} \hline
{\it Representation} & $(-1)^k \eta_{k}$ & $(-1)^k \eta'_{k}$ &
 $n_{\bar{d}}$ & $n_l$ & $n_{\bar{u}}$ & $n_{\bar{e}}$ & $n_q$ & $n_{\bar{\nu}}$ \\ \hline\hline
$[8,1]$ & $+1$ & $+1$ & 1 & 0 & 0 & 0 & 0 & 0 \\
 & $+1$ & $-1$ & 0 & 1 & 0 & 0 & 0 & 3 \\ \hline
$[8,2]$ & $+1$ & $+1$ & 0 & 3 & 1 & 1 & 0 & 3 \\
 & $+1$ & $-1$ & 3 & 0 & 0 & 0 & 1 & 0 \\ \hline
$[8,3]$ & $+1$ & $+1$ & 3 & 0 & 1 & 1 & 3 & 0 \\
 & $+1$ & $-1$ & 0 & 3 & 3 & 3 & 1 & 1 \\ \hline
$[8,4]$ & $+1$ & $+1$ & 1 & 1 & 3 & 3 & 3 & 0 \\
 & $+1$ & $-1$ & 1 & 1 & 3 & 3 & 3 & 0 \\ \hline
$[8,5]$ & $+1$ & $+1$ & 0 & 3 & 3 & 3 & 1 & 1 \\
 & $+1$ & $-1$ & 3 & 0 & 1 & 1 & 3 & 0 \\ \hline
$[8,6]$ & $+1$ & $+1$ & 3 & 0 & 0 & 0 & 1 & 0 \\
 & $+1$ & $-1$ & 0 & 3 & 1 & 1 & 0 & 3 \\ \hline
$[8,7]$ & $+1$ & $+1$ & 0 & 1 & 0 & 0 & 0 & 3 \\
 & $+1$ & $-1$ & 1 & 0 & 0 & 0 & 0 & 0 \\ \hline
\end{tabular}
\end{center}
\end{table}

\begin{table}
\caption{The flavor number of each chiral fermion from $[9, k]$ for $r = 3$.}
\label{t6}
\begin{center}
\begin{tabular}{c|c|c|c|c|c|c|c|c} \hline
{\it Representation} & $(-1)^k \eta_{k}$ & $(-1)^k \eta'_{k}$ &
 $n_{\bar{d}}$ & $n_l$ & $n_{\bar{u}}$ & $n_{\bar{e}}$ & $n_q$ & $n_{\bar{\nu}}$ \\ \hline\hline
$[9,1]$ & $+1$ & $+1$ & 1 & 0 & 0 & 0 & 0 & 1 \\
 & $+1$ & $-1$ & 0 & 1 & 0 & 0 & 0 & 3 \\ \hline
$[9,2]$ & $+1$ & $+1$ & 1 & 3 & 1 & 1 & 0 & 3 \\
 & $+1$ & $-1$ & 3 & 1 & 0 & 0 & 1 & 3 \\ \hline
$[9,3]$ & $+1$ & $+1$ & 3 & 3 & 2 & 2 & 3 & 3 \\
 & $+1$ & $-1$ & 3 & 3 & 3 & 3 & 2 & 1 \\ \hline
$[9,4]$ & $+1$ & $+1$ & 4 & 1 & 4 & 4 & 6 & 0 \\
 & $+1$ & $-1$ & 1 & 4 & 6 & 6 & 4 & 1 \\ \hline
$[9,5]$ & $+1$ & $+1$ & 1 & 4 & 6 & 6 & 4 & 1 \\
 & $+1$ & $-1$ & 4 & 1 & 4 & 4 & 6 & 0 \\ \hline
$[9,6]$ & $+1$ & $+1$ & 3 & 3 & 3 & 3 & 2 & 1 \\
 & $+1$ & $-1$ & 3 & 3 & 2 & 2 & 3 & 3 \\ \hline
$[9,7]$ & $+1$ & $+1$ & 3 & 1 & 0 & 0 & 1 & 3 \\
 & $+1$ & $-1$ & 1 & 3 & 1 & 1 & 0 & 3 \\ \hline
$[9,8]$ & $+1$ & $+1$ & 0 & 1 & 0 & 0 & 0 & 3 \\
 & $+1$ & $-1$ & 1 & 0 & 0 & 0 & 0 & 1 \\ \hline
\end{tabular}
\end{center}
\end{table}

{}From Table \ref{t5} and \ref{t6}, we find that the flavor numbers from $[N, k]$ 
with the intrinsic $Z_2$ parity $((-1)^k \eta_{k}, (-1)^k \eta'_{k}) = (+1, \pm 1)$
are equal to those from $[N, N-k]$ with $((-1)^{N-k} \eta_{N-k},$ $(-1)^{N-k} \eta'_{N-k})$ $= (+1, \mp 1)$ for $N=8, 9$ and $r=3$.
This kind of relation is generalized that, for arbitrary $N (\ge 6)$ and $r$, the flavor numbers from $[N, k]$ 
with $((-1)^k \eta_{k},$ $(-1)^k \eta'_{k})$ $= (a, b)$
equal to those from $[N, N-k]$ with $((-1)^{N-k} \eta_{N-k}$, $(-1)^{N-k} \eta'_{N-k})$ $= (a, -b)$ if $r$ is odd
and the flavor numbers from $[N, k]$ 
with $((-1)^k \eta_{k},$ $(-1)^k \eta'_{k})$ $= (a, b)$
equal to those from $[N, N-k]$ with $((-1)^{N-k} \eta_{N-k},$ $(-1)^{N-k} \eta'_{N-k})$ $= (a, b)$ if $r$ is even.
The proof goes as follows.
The representation $[N,N-k]$ is decomposed into a sum of multiplets as
\begin{eqnarray}
[N, N-k] = \sum_{3-l_1 =0}^{N-k} \sum_{2-l_2 = 0}^{N-k-3+l_1} \sum_{r-l_3 = 0}^{N-k-5+l_1+l_2}  
\left({}_{p}C_{3-l_1}, {}_{q}C_{2-l_2}, {}_{r}C_{r-l_3}, {}_{s}C_{s-l_4}\right) .
\label{NN-k}
\end{eqnarray}
There is a one-to-one correspondence among each multiplet in $[N, N-k]$ and $[N, k]$, e.g.,
the right-handed Wely fermion with $\left({}_{p}C_{3-l_1}, {}_{q}C_{2-l_2}, {}_{r}C_{r-l_3}, {}_{s}C_{s-l_4}\right)_R$ 
corresponds to the left-handed one with $\left({}_{p}C_{l_1}, {}_{q}C_{l_2}, {}_{r}C_{l_3}, {}_{s}C_{l_4}\right)_L$ 
by the charge conjugation.
The $Z_2$ parity assignment of $\left({}_{p}C_{3-l_1}, {}_{q}C_{2-l_2}, {}_{r}C_{r-l_3}, {}_{s}C_{s-l_4}\right)_R$ is given by
\begin{eqnarray}
&~& \mathcal{P}_0 = -(-1)^{r-l_3+s-l_4} \eta_{N-k} = (-1)^{l_1+l_2} (-1)^{N-k} \eta_{N-k} , 
\label{Z2-N-k}\\
&~& \mathcal{P}_1 = -(-1)^{2-l_2+s-l_4} \eta'_{N-k} = (-1)^r (-1)^{l_1+l_3} (-1)^{N-k} \eta'_{N-k} .
\label{Z'2-N-k}
\end{eqnarray}
By comparison of (\ref{Z2-N-k}) and (\ref{Z'2-N-k}) with (\ref{Z2L}), we arrive the relation for the flavor numbers 
stemming from different representations.

There are many possibilities that zero modes from a bulk multiplet and a few brane fields compose three families.
For example, we find that a single bulk field $[9, 3]$ and two brane fields with the same gauge quantum number 
as $({u}_R)^c$ and $({e}_R)^c$ make up three families.

\subsection{Validity of our analysis}
\label{validity}
Finally we discuss the validity of our analysis.
For our setup, each $SU(2)$ sub-block connecting $p$ and $s$ parts, or $q$ and $r$ parts, can in principle develop a Wilson line to further break symmetry by the Hosotani mechanism, which we explain in this subsection.

In gauge theory, physics should not depend on the gauge choice and we are always free to choose a gauge.
The two set of BCs are equivalent if they are related to each other by a large gauge transformation defined in the covering space~$R^1$.
For example using the $SU(2)$ large gauge transformation with the gauge function $\Omega(y) = \exp\{i(y/2R)\tau_2\}$, 
we find the following equivalence
\begin{eqnarray}
(P_0 = \tau_3 , P_1 = \tau_3) \sim (P_0 = \tau_3 , P_1 = -\tau_3) ,
\label{equ}
\end{eqnarray}
where $\tau_i$s are Pauli matrices.
{}From the relation (\ref{equ}), we can derive the following equivalence relations~\cite{HH&K}
\begin{align}
[p; q, r; s] 
	&\sim [p-1; q+1, r+1; s-1] ,~~ (\mbox{for}~~ p, s \ge 1) ,
\nonumber \\
	&\sim [p+1; q-1, r-1; s+1] ,~~ (\mbox{for}~~ q, r \ge 1) .
\label{equ-SUN}
\end{align}
The symmetry of the BCs in one theory differs from that in the other, but if the two theories are connected by the BCs-changing gauge 
transformation, they are equivalent when we neglect the Wilson line.
This equivalence is gauranteed by the Hosotani mechanism~\cite{H,HHH&K}.
We explain it briefly.

First consider the system described by BCs $P_0$ and $P_1$.
Generally an effective potential for $A_y$ is generated at quantum level so that $A_y$ acquires a vacuum expectation value and that the Wilson line
\begin{eqnarray}
 W \equiv \mathcal{P}\exp\left(ig \int_{-\pi R}^{\pi R} A_y dy\right)
\label{W}
\end{eqnarray}
takes non-zero value at the vacuum, where $g$ is the gauge coupling constant and $\mathcal{P}$ is the path ordering.
The physical gauge invariant degrees of freedom are the eigenvalues of the matrix $WU$, where $U = P_1 P_0$.  
The symmetry of the system is individualized by $(P_0, P_1; W)$.

Next we perform a large gauge transformation that is continuous and single valuedf 
in the covering space $R^1$ in order to eliminate the Wilson line $W$.
Then the BCs change into different ones $(P_0^\text{sym}, P_1^\text{sym}; I)$.
We note that $(P_0, P_1; W)$ and $(P_0^\text{sym}, P_1^\text{sym}; I)$ represent the physically equivalent system.
In this gauge with vanishing Wilson line, the physical symmetry is spanned by the generators that commute with $P_0^\text{sym}$ and $P_1^\text{sym}$.
These matrices are not necessalily diagonal, but one of them, say $P_0^\text{sym}$, 
can be diagonalized through a global gauge transformation.
Then $P_1^\text{sym}$ is not diagonal in general.
After reshuffling rows and columns and performing global unitary transformations, $P_0^\text{sym}$ and  $P_1^\text{sym}$ take the following standard form
\begin{align}
P_0^\text{sym}
	&= \blockdiag(I_{p'},I_{q'},-I_{r'},-I_{s'},\overbrace{\tau_3,\dots,\tau_3}^{n'}),\nonumber\\
P_1^\text{sym}
	&= \blockdiag(I_{p'},-I_{q'},I_{r'},-I_{s'},P_1^{(1)},\dots,P_1^{(n')}),
		\label{P1sym}
\end{align}
where 
	`$\blockdiag$' stands for the block diagonal matrix,
	$I_p$ is the $p\times p$ unit matrix,
	$N = N' + 2n'$, 
	$N' = p' + q' + r' + s'$, and 
	$P_1^{(a)} = e^{-2\pi i \alpha_a \tau_2} \tau_3$
		($a = 1, \dots, n'$) with non-integer $\alpha_a$.
(If $\alpha_a$ is an integer, that sub block can be reshuffled into the diagonal entry.)
We refer to the transformation group regarding $P_1^{(a)}$s as `twisted $SU(2)$'.
The physical symmetry is $SU(p') \times SU(q') \times SU(r') \times SU(s') \times U(1)^{2n'+\nu}$.
This is the most general BCs in $SU(N)$ gauge theory on the $S^1/Z_2$.

Now we show that no zero mode survives for the doublet of twisted $SU(2)$ after compactification with the BCs (\ref{P1sym}).
The mode expansion of a bulk field depends on representations under the twisted $SU(2)$s.
For a field $\Phi$ that is doublet under a twisted $SU(2)$ with the $Z_2$ parity $(P_0, P_1)=(\tau_3, e^{-2 \pi i \alpha \tau_2} \tau_3)$, 
the orbifold BCs are
\begin{align}
\Phi(x, -y) 
	&= \tau_3 \Phi(x, y) , \label{Phi-P0}\\
\Phi(x, y+2\pi R) 
	&= e^{-2 \pi i \alpha \tau_2} \Phi(x, y) = 
\left(
\begin{array}{cc}
\cos 2\pi \alpha & -\sin 2\pi \alpha \\
\sin 2\pi \alpha & \cos 2 \pi \alpha
\end{array}
\right) \Phi(x, y),
\label{Phi-U}
\end{align}
and the Fourier expansion is given by 
\begin{eqnarray}
\left(
\begin{array}{c}
\phi^1(x, y)  \\
\phi^2(x, y)  
\end{array}
\right) = \frac{1}{\sqrt{\pi R}} \sum_{n = \infty}^{\infty} \phi_n(x) 
\left(
\begin{array}{c}
\cos \frac{(n+\alpha)y}{R}  \\
\sin \frac{(n+\alpha)y}{R}  
\end{array}
\right) .
\label{Phi-expansion}
\end{eqnarray}
From the expansion (\ref{Phi-expansion}), we can see that the $SU(2)$ doublet have no zero mode for non-integer~$\alpha$.

The decomposition of $[N, k]$ contains singlets and doublets of twisted $SU(2)$s.
When we start with a representation $[N, k]$, zero modes must be necessarily singlets under all the twisted $SU(2)$s.
Because there are two kinds of twisted $SU(2)$ singlets ${}_{2}C_{0}$ and ${}_{2}C_{2}$ for each $SU(2)$, 
the representation $[N, k]$ is reduced to the sum of representations $\sum_{l=0}^{n'} {}_{n'}C_l [N', k-2l]$ 
in the presence of twisted $SU(2)$s, where $l$ is the number of ${}_{2}C_{2}$ in each term.
In this case, our analysis has to be repeated with $\sum_{l=0}^{n'} {}_{n'}C_l [N', k-2l]$, where the $Z_2$ parity assignment for each term is given by
\begin{align}
P_0^\text{sym} 
	&= \diag(\overbrace{+1, \dots, +1, +1, \dots, +1, -1, \dots, -1, -1, \dots, -1}^{N'}) , 
\label{P'0} \\
P_1^\text{sym} 
	&= \diag(\underbrace{+1, \dots, +1}_{p'}, \underbrace{-1, \dots, -1}_{q'}, \underbrace{+1, \dots, +1}_{r'}, 
 \underbrace{-1, \dots, -1}_{s'}) .
\label{P'1}
\end{align}

Finally we list the flavor number of SM fermions in the case $p' = 3$, $q' = 2$ 
and $(-1)^k \eta_{k} = (-1)^k \eta'_{k} = +1$ ,
\begin{align}
n_{\bar{d},k}^{(++)} 
	&= \sum_{l=0}^{n'} \sum_{i = 1, 4} \sum_{l_3 = 0, 2, \dots} {}_{n'}C_l \cdot {}_{r'}C_{l_3} \cdot {}_{N'-5-r'}C_{k-2l-i-l_3} ,
\label{nd-gen}\\
n_{l,k}^{(++)} 
	&= \sum_{l=0}^{n'} \sum_{i = 1, 4} \sum_{l_3 = 1, 3, \dots} {}_{n'}C_l \cdot {}_{r'}C_{l_3} \cdot {}_{N'-5-r'}C_{k-2l-i-l_3} ,
\label{nl-gen}\\
n_{\bar{e},k}^{(++)} = n_{\bar{u},k}^{(++)}  
	&= \sum_{l=0}^{n'} \sum_{i = 2, 3} \sum_{l_3 = 0, 2, \dots} {}_{n'}C_l \cdot {}_{r'}C_{l_3} \cdot 
 {}_{N'-5-r'}C_{k-2l-i-l_3} ,
\label{ne-gen}\\
n_{q,k}^{(++)} 
	&= \sum_{l=0}^{n'} \sum_{i = 2, 3} \sum_{l_3 = 1, 3, \dots} {}_{n'}C_l \cdot {}_{r'}C_{l_3} \cdot {}_{N'-5-r'}C_{k-2l-i-l_3} ,
\label{nq-gen}\\
n_{\bar{\nu},k}^{(++)} 
	&= \sum_{l=0}^{n'} \sum_{i = 0, 5} \sum_{l_3 = 0, 2, \dots} {}_{n'}C_l \cdot {}_{r'}C_{l_3} \cdot {}_{N'-5-r'}C_{k-2l-i-l_3} .
\label{nnu-gen}
\end{align}
Using Eqs.~(\ref{nd-gen})--(\ref{nnu-gen}), we can obtain the flavor numbers, e.g.,
$n_{\bar{d}} = n_{\bar{u}} = n_{\bar{e}} = n_{q} = 3$, $n_{l} = n_{\bar{\nu}} = 1$ from $[N, k]=[12, 5]$ 
in the case that $N'=6$, $n'=3$, and $\eta_{5} = \eta'_{5} = -1$.

We note that still it is necessary to check if the wanted symmetry breaking pattern occurs dynamically 
via the Hosotani mechanism for a given matter content, which we leave for future study.

\section{Conclusion and discussions}

We have presented the idea of the complete family unification in higher-dimensional space-time
and found that three families in $SU(5)$ grand unified theory can be derived from a single bulk multiplet 
of $SU(N)$ $(N \ge 9)$ in the framework of gauge theory on the orbifold $S^1/Z_2$.
In the case of the direct orbifold breaking down to the standard model gauge group,
there are models in which bulk fields from a single multiplet and a few brane fields compose three families,
e.g.\ a single bulk field $[9, 3]$ and two brane fields make up three families. 
The flavor numbers of the SM fermions have been also written down in the case 
that a totally antisymmetric representation $[N, k]$ of $SU(N)$ has the most general BCs of $S^1/Z_2$ including twisted $SU(2)$s. 
It would also be interesting to study the case that bulk fields form an irreducible representation with mixed symmetry.

There are several open questions towards more realistic model-building which are left for future work.

In our setup, all the unwanted matter degrees of freedom are successfully made massive thanks to the orbifolding. 
However, all the $SU(r')\times SU(s')$ gauge fields remain massless even after the symmetry breaking due to the Hosotani mechanism. 
In most cases, this kind of non-abelian subgroup of $SU(N)$ plays the role of family symmetry.
These massless degrees of freedom must be made massive by further breaking of the family symmetry.
Here we point out that the brane fields can be key to the solutions.
Most models have chiral anomalies at the four-dimensional boundaries and
we have a choice to introduce appropriate brane fields to cancel these anomalies.
Further, some brane fields can play a role of Higgs fields for the breakdown of extra gauge symmetries including the non-abelian gauge symmetries.
As a result, extra massless fields including the family gauge bosons can be made massive.

Can the gauge coupling unification be successfully achieved?
If the MSSM particle contents only remain in the low-energy spectrum around and below the TeV scale 
and a big desert exists after the breakdown of extra gauge symmetries, 
an ordinary grand unification scenario can be realized up to the threshold corrections from the Kaluza-Klein modes 
and from the brane contributions from non-unified gauge kinetic terms.

Another problem is whether or not the realistic fermion mass spectrum and the generation mixings are successfully achieved.
Fermion mass hierarchy and generation mixings can also occur through the Nielsen-Froggatt mechanism 
concerning to extra gauge symmetries and the suppression of brane-localized Yukawa coupling constants 
among brane weak Higgs doublets and bulk matters with the volume suppression factor.

In general, there appear $D$-term contributions to scalar masses in SUSY models 
after the breakdown of such extra gauge symmetries~\cite{D,H&K,KM&Y}.
The $D$-term contribution concerning family symmetry spoils the degeneracy of sfermion masses, and
can induce large flavor changing neutral current (FCNC) processes that is incompatible with experimental data.
This would give a tight constraint on model-building and it would be an interesting challenge 
to construct a realistic model from the orbifold GUT with a single bulk matter multiplet.
 
The orbifold GUT is more naturally realized in warped space, see e.g.~\cite{Nomura:2004zs} for a review. 
It has been shown that the argument for the Hosotani mechanism presented in Section~\ref{validity} 
can be followed in a parallel manner in warped space too~\cite{Oda:2004rm}. 
Therefore, it would be interesting to look for a more realistic version of our model in warped space.
In such a scenario, one of the scales concerning the orbifold breaking with $P_0$ and $P_1$,
as well as the dynamical symmetry breaking scale of the Hosotani mechanism, would be exponentially suppressed from the GUT breaking energy scale.

It would be interesting to study cosmological implications of the class of models presented in this paper, 
see e.g.~\cite{Khlopov:1999rs} and references therein for useful articles toward this direction.

\bigskip

\noindent \textit{Note added:} After completion of this work, there appeared a paper treating the related subject~\cite{Gogoladze:2007nb}.

\section*{Acknowledgements}
This work was supported in part by Scientific Grants from the Ministry of Education and Science, 
Grant No.13135217, Grant No.18204024, Grant No.18540259 (Y.K.) and by the Special Postdoctoral Researchers Program at RIKEN (K.O.).

\end{document}